\documentclass[usegraphicx,usedcolumn,usenatbib]{mn2e}
\usepackage{graphics,longtable,natbib}

\title{Physical parameters of the O6.5V+B1V eclipsing binary system LS~1135}
\author[E.~Fern\'andez Laj\'us et al.]
{
E.~Fern\'andez Laj\'us 
\thanks{Fellow of CONICET, Argentina.}
and V.S.~Niemela 
\thanks{Member of the Carrera del Investigador Cient\'{\i}fico, CIC-BA, Argentina.}
\\
Facultad de Ciencias Astron\'omicas y Geof\'{\i}sicas, Universidad Nacional de La Plata, Argentina. \\
}

\date{}

\pagerange{\pageref{firstpage}--\pageref{lastpage}}
\pubyear{2005}

\begin{document}

\label{firstpage}

\maketitle

\begin{abstract}
ASAS photometric observations of LS~1135, an O--type SB1 binary system with
an orbital period of 2.7 days, show that the system is also eclipsing.
This prompted us to re--examine the spectra used in the previously published
spectroscopic orbit. Our new analysis of the spectra obtained
near quadratures, reveal the
presence of faint lines of the secondary component. We present for the first
time a double--lined radial velocity orbit and values of physical 
parameters of this binary system. These values were obtained by analyzing
ASAS photometry jointly with the radial velocities of both components performing
a numerical model of this binary based on the Wilson--Devinney method.
We obtained an orbital inclination $i \sim 68^{\circ}.5$. With this value of
the inclination we deduced masses  $M_1 \sim 30 \pm 1 M_{\odot}$
and $M_2 \sim 9 \pm 1 M_{\odot}$;  and radii $R_1 \sim 12 \pm 1 R_{\odot}$ 
 and $R_2 \sim 5 \pm 1 R_{\odot}$ for primary and secondary components, 
respectively. Both components are well inside their respective Roche lobes.
Fixing the $T_{eff}$ of the primary to the value corresponding to its 
spectral type (O6.5V), the $T_{eff}$ obtained for the secondary component
corresponds approximately to a spectral type of B1V.
The mass ratio $M_2 /M_1 \sim 0.3$ is among the lowest known values
for spectroscopic binaries with O--type components.
\\
\end{abstract}


\section{Introduction}
The star LS 1135 ($\alpha_{2000.0} = 08^{h} 43^{m} 50^{s}$,{}
 $\delta_{2000.0}$ = -$46^\circ $ 07' 09"), a member of the OB association 
Bochum 7 (Vela OB 3), was discovered by
Corti et al. (2003, hereafter CNM) to be a single-lined binary (SB1)
system with an orbital period of 2.7532 days. The spectrum was classified 
as O6.5V ((f)). The rather high amplitude of the radial velocity variations and
the lack of detected spectral lines of the secondary component led CNM
to suggest that the secondary might be an early B type star.

Magnitudes and colors for LS~1135 from photoelectric photometry have been
published by Moffat \& Vogt (1975) as $V = 10.88, B-V = 0.4, U-B = -0.68$,
and by Drilling (1991) as $V = 10.88, B-V = 0.38, U-B = -0.66$.
LS~1135  was found to be a photometrically variable star in 
the ``All Sky Automated Survey'' (ASAS) (cf. Pojma\'nski 2003).
It is catalogued as an eclipsing system, and a period of 2.7532 days 
was found independently in the photometric data, in
perfect agreement with the period obtained from the radial velocities by
CNM.
The fact that LS~1135 is an eclipsing binary, motivated us to  search
for signatures of the secondary component by reinspecting the spectra used by CNM.
Here we report the discovery of the spectral lines of the secondary, and
calculate for the first time a double--lined radial velocity orbit and a 
set of values of the main physical parameters of the components
of LS~1135. These are based on a simultaneous analysis of the 
photometric light curve and radial velocity variations of both components 
by means of the
Wilson-De\-vinney (W--D) Code (Wilson \& De\-vinney 1971, Wilson 1990, 
Wilson \& Van Hamme 2004).

\begin{figure}
\includegraphics[width=240pt]{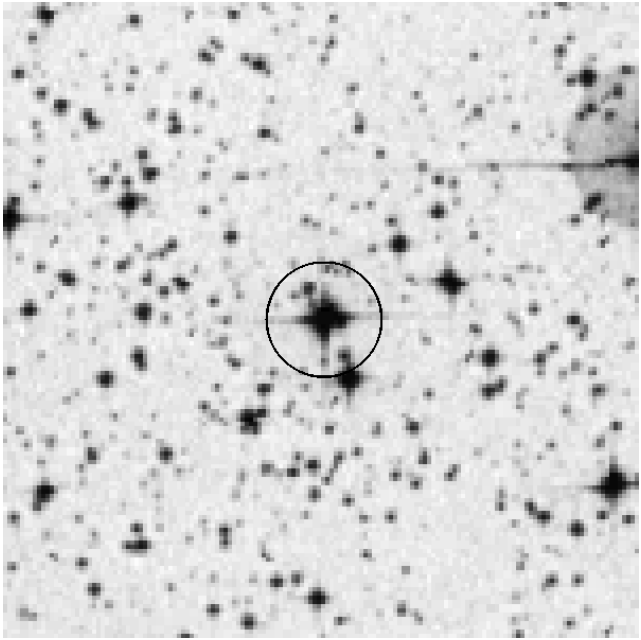}
\caption{A 4' x 4' image of the Digitized Sky Survey, centered in LS~1135.
         The circle represents the ASAS 3-pixels (= 46".5) photometric
	 aperture. (North is up and East to the left).}
        \label{DSS1}
\end{figure}

\section{Observations}
\subsection{The ASAS V light curve of LS 1135}
ASAS (cf. Pojma\'nski 2002) has been monitoring photometrically a large 
part of the sky down to stellar magnitude $V \sim 14^{mag}$.
The ASAS instrument was upgraded in late 2000, being then called ASAS-3
(Pojma\'nski, 2001). The current configuration obtains
images with a wide-field (8$.\!{\!^\circ}$8 x 8$.\!{\!^\circ}$8) CCD
camera, using a standard $V$ filter.
The scale ($\sim 15"/pixel$) of the system results in sub-sampled
images, which are suitable to be analyzed through aperture photometry.
Figure~\ref{DSS1} shows a 4' x 4' region of the sky around LS~1135.

\begin{figure*} 
\includegraphics[width=180pt, angle=270]{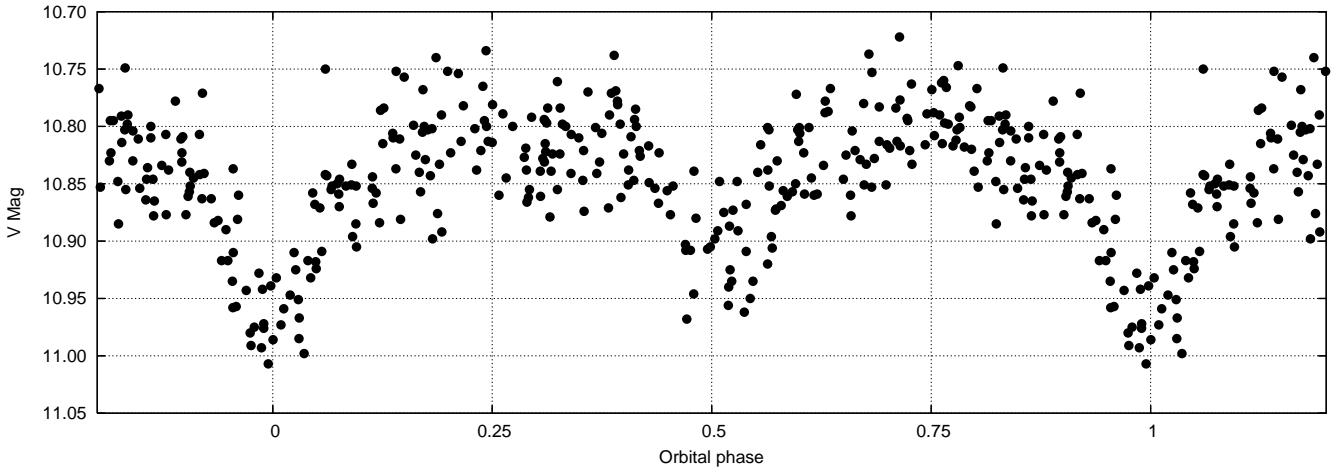}
\caption{The ASAS-3 V light curve of LS 1135. Magnitudes are those measured
using a 46".5 diameter aperture. The data correspond to observations
adquired between 2000, November, and 2005, July.}
        \label{ASASVLC} 
\end{figure*}

LS~1135, catalogued as ASAS 084350-4607.2, is classified as an eclipsing 
binary of type ESD/ED (semi-detached/detached) in the ASAS-3 Catalog
of Variable Stars, available in the Internet (cf. ASAS Web Site).
We have retrieved from the ASAS-3 database the $V$ photometric data of LS~1135
observed between 2000, November, and 2005, July. 
The light curve is shown in Figure~\ref{ASASVLC}.
The data in this figure correspond to $V$ values measured with the 3 pixel
(= 46".5) aperture. We selected 313 data values 
corresponding to this aperture because they show the light curve with the 
lowest dispersion of data points, and avoid the introduction of third light 
due to the star located 27'' to the S-W of LS 1135 (see Fig.~\ref{DSS1}). 

We note that there are several much fainter stars included in the photometric
aperture. Taking into account that according to the 
``The Tycho-2 Catalogue of the 2.5 Million Brightest Stars'' (H\o g et al. 2000), 
the magnitude difference between LS 1135 and the neighbour at 27'' to the S-W
is $\sim$ 2.3 magnitudes, the other stars included in the aperture appear to be more
than 5 magnitudes fainter than LS 1135. Thus their contribution to the total
light would be less than 1\%, which has been ignored in our analysis.

The zero point calibration used by ASAS photometry is based on the Hipparcos 
photometry (Perryman et al. 1997).

\subsection{Spectroscopy}
Photographic and digital spectra in the blue spectral region, 
as previously described by CNM, were used in our study of LS 1135.
We have re--examined those spectra observed near the quadratures of the orbital
motion in a search for spectral signatures of the secondary component.
We found that  some of the He{\sc i}  absorption lines appear double near the
maximum radial velocity separation of the components. We could measure the
radial velocities from these lines due to the secondary component in ten spectra. 
The values of these radial velocities were obtained fitting gaussians to the 
spectral lines within the IRAF  \footnote{IRAF is distributed by the NOAO, 
operated by the AURA, Inc, under cooperative agreement with the NSF, USA.}  
routine {\it splot}, and are  tabulated in Table~\ref{tRV}
according the Heliocentric Julian Date (HJD) of observations.
The instrumental configuration is given in column 3, following
the notation from Table 1 in CNM.
Typical errors in the determination of the radial velocities of the faint 
lines of the secondary component are rather large, $\sim$\,40~km~$s^{-1}$.

\begin{table}
\center
\caption{Observed heliocentric radial velocities of the spectral lines
        of the secondary component of LS 1135. IC refers to the
        instrumental configuration in CNM.}
\label{tRV}
\begin{tabular}{c c c}
\noalign{\smallskip}\hline\noalign{\smallskip}
HJD          & RV               & IC \\
2\,400\,000- & (km~s$^{-1}$)         \\
\noalign{\smallskip}\hline\noalign{\smallskip}
45508.463 & -360 & I   \\
45511.453 & -302 & I   \\ 
50508.549 & -276 & II  \\
50538.586 & -277 & II  \\
50541.567 & -289 & II  \\
50845.620 &  440 & II  \\
50848.609 &  350 & II  \\
50859.600 &  406 & III \\
50860.652 & -287 & III \\ 
50965.447 & -307 & III \\
\noalign{\smallskip}\hline\noalign{\smallskip}
\\
\end{tabular}
\end{table}

\section{The radial velocity orbit}
As a first step to estimate values of the physical parameters 
of the components of the binary system LS~1135, we determined a double--lined 
radial velocity orbit for the binary. 
For this we adjusted circular orbits to the complete set of 
radial velocities i.e. those published by CNM for the primary, and the values
listed in Table~\ref{tRV} for the secondary. Because the radial velocities of
the secondary component have considerably larger errors, they were assigned
lower weight.  For the determination of the orbital elements we used an 
improved version of the code originally published by Bertiau \& Grobben (1969).
Eccentricity was fixed to 0.0 since both minima in the light curve appear 
symmetrically separated. 
We also calculated an orbital solution leaving the eccentricity as a free 
parameter. This solution  gave a very small value of the eccentricity,
comparable with the error of the circular orbit. We have therefore
adopted $e = 0$ for further analysis in this paper.

The initial value for the period was set to 2.7532 days from the previously
published single--lined radial velocity orbit.
 The value obtained for the double--lined radial velocity orbit resulted  
in almost the same value, namely 2.753205 $\pm ~10^{-5}$ days. 
The values of the orbital parameters are presented in Table~\ref{tBert}.
These values are to be considered as preliminary estimates, awaiting more
accurate radial velocities for the secondary component from high resolution
spectra with higher signal-to-noise ratio.

\begin{table}
\center
\caption{Circular orbital elements from radial velocities of both components of LS~1135}
\label{tBert}
\begin{tabular}{l c c}
\noalign{\smallskip}\hline\noalign{\smallskip}
Element          		&  Primary     		&  Secondary \\
\noalign{\smallskip}\hline\noalign{\smallskip}
a sin i [$R_{\odot}$]   	& 6.15 $\pm$ 0.13	& 19.98 $\pm$ 0.13 \\
K [km~s$^{-1}$]			& 114 $\pm$ 3       & 370 $\pm$ 13 \\
M $sin^3 i$ [$M_{\odot}$]	& 24.6 $\pm$ 3.3 	& 7.6 $\pm$ 1.2 \\
$M_2/M_1$       		& \multicolumn{2}{c}{0.31 $\pm$ 0.02} \\
$T_{RV_{Max}}$ [HJD] 		& \multicolumn{2}{c}{2445508.5 $\pm$ 0.1} \\     
Period [days] 			& \multicolumn{2}{c}{2.753205 $\pm ~10^{-5}$} \\
$V_{\gamma}$ [km~s$^{-1}$] 	& \multicolumn{2}{c}{65 $\pm$ 2} \\
\noalign{\smallskip}\hline\noalign{\smallskip}
\\
\end{tabular}
\end{table}

\section{W--D eclipsing binary model for the light and radial velocity 
curves of LS 1135}

In order to obtain the absolute values of the physical parameters of 
the binary components, 
we need to determine the orbital inclination $i$ of the system.
To this end, we adjusted a numerical eclipsing binary model to the 
observations, using the W--D Code. 
We used the PHOEBE package tool (Pr\v sa \& Zwitter, 2005) 
for the light curve (LC) and differential corrections (DC) fits.
ASAS photometric data and radial velocities of both components of the binary
system were used to adjust the model.

The code was set in Mode 2 for detached binaries with no constraints on the 
potentials (except for the luminosity of the secondary component).
The simplest considerations were applied for the emission parameters of the 
stars in the model, i.e. stars were considered as black bodies, 
approximate reflection model (MREF=1) was adopted,
and no third light or spots were included. Gravity darkening exponents 
$g_{1} = g_{2} = 1$ and bolometric albedos $Alb_{1} = Alb_{2} = 1$ were set 
for radiative envelopes.
We used the square root limb darkening law. Bolometric and $V$ band limb 
darkening coefficients were taken from Van Hamme (1993).
We adopted the mass ratio $q = M_{2}/M_{1} = 0.31$ from our radial velocity orbit. 
The temperature for the primary component was set to the corresponding value
from the Spectral Type-$T_{eff}$ calibration tables published by Martins et al. 
(2005). $T_1 $ was fixed in this value and $T_2$ was computed
with the model fit.
We considered that the system has a circular orbit and that both components 
rotate syncronously ($F_{1} = F_{2} = 1$). 
The radial velocity of the centre of mass of the binary system was fixed
in the value obtained from the radial velocity orbit.

\begin{figure*}
	\includegraphics[width=180pt, angle=270]{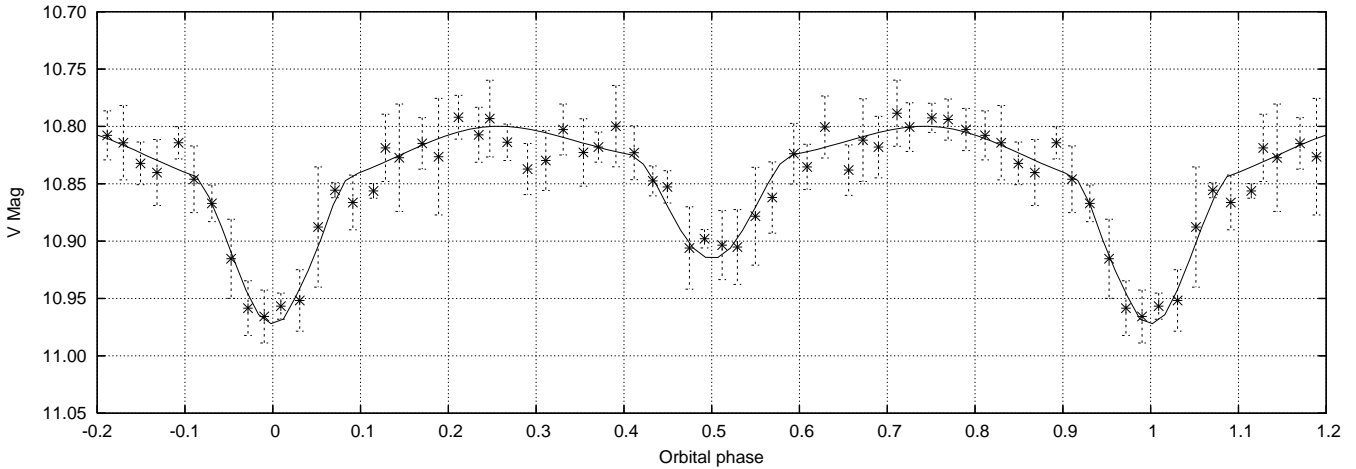}
	\caption{Normal points of the ASAS $V$ light curve of LS~1135
phased with the binary period of 2.7532~days. 
	The continuous line represents our best fit W-D model.}
	\label{CL}
\end{figure*}

\begin{figure}
	\includegraphics[width=200pt, angle=270]{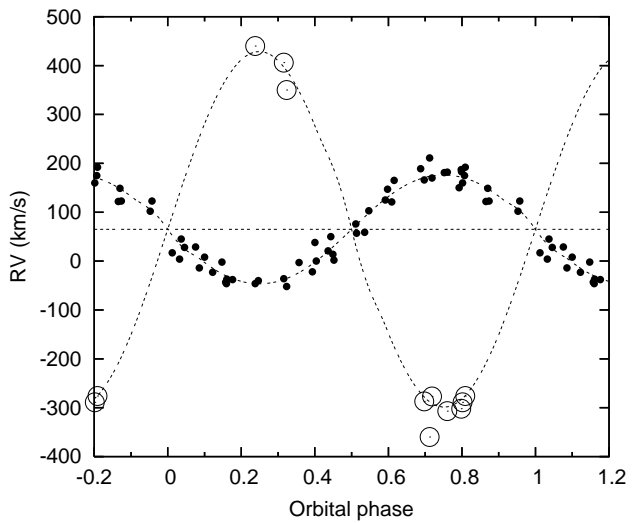}
	\caption{Observed radial velocities of both components of LS~1135
phased with the binary period of 2.7532~days.
	The curves represent the best fit W-D model.}
	\label{VR}
\end{figure}

Photometric data were weighted from the original magnitude errors provided 
by the ASAS database taking the weight as $w \propto 1/m^2_{err}$, 
in such a way that the 
observations with the lowest magnitude error had $weight = 1$. 
We then generated  synthetic light and radial 
velocity curves by means of the $LC$ program of the W-D code, and adjusted 
them to the observations. 
In order to get a better delineated light curve, we proceeded to calculate 
normal points of the ASAS photometric data taking 0.02 phase bins. 
A weighted average of the magnitudes 
was computed for every bin, using the individual errors provided by ASAS 
database. A typical $rms \sim  0.024~mag$ is present in each mean point
for the aperture considered in this paper.
The results of our best fitting model are depicted in Figures~\ref{CL} and 
\ref{VR} together with the normal points of the ASAS light curve of LS~1135,
and the observed radial velocities of both components, respectively.
The physical parameters of the system are presented in Table~\ref{tWD}. 

\begin{table}
\center
\caption{Preliminary physical parameters of the binary components of LS 1135 
from the best fitting WD Model. $R_L$ stands for effective radius of the
Roche lobe (cf. Eggleton 1983).}
\label{tWD}
\begin{tabular}{c c c}
\noalign{\smallskip}\hline\noalign{\smallskip}
	Parameter  &    Primary   &  Secondary   \\
	\noalign{\smallskip}\hline\noalign{\smallskip}
	Period [$days$]     	&\multicolumn{2}{c}{$2.753205 \pm 2\times10^{-5}$}\\
	$T_0$ [HJD]		&\multicolumn{2}{c}{$2451871.91 \pm 1\times10^{-2}$}\\
	V$\gamma$ [$km s^{-1}$] &\multicolumn{2}{c}{$65 ^{\ast}$}\\
        $M_{2}$/$M_{1}$     	&\multicolumn{2}{c}{0.31$^\ast$}\\
	$i$ [$^\circ$]      	&\multicolumn{2}{c}{68.5 $\pm$ 1}\\
	a [$R_{\odot}$]     	&\multicolumn{2}{c}{28.1 $\pm$ 0.2}\\
	M [$M_{\odot}$]     	& 30 $\pm$ 1	  & 9 $\pm$ 1\\
	$R_{mean}$ [$R_{\odot}$]& 12 $\pm$ 1   &  5 $\pm$ 1 \\
	Teff [$^{\circ}\!K$]   	& 37870 $^a$ & 25500 $\pm$ 500 \\
	Mbol                	&-8.8 $\pm$ 0.1  & -5.3 $\pm$ 0.1 \\
	$L_{2}/L_{1}$   	&\multicolumn{2}{c}{0.11 $\pm$ 0.05}\\
	Log g [cgs]         	&3.76 $\pm$ 0.05  & 3.96 $\pm $0.05 \\
	$R_L$ [$R_{\odot}$]	&13.6 $\pm$ 0.1 & 7.95 $\pm$ 0.1 \\
	\noalign{\smallskip}\hline\noalign{\smallskip}
	\multicolumn{3}{l}{$\ast$: Fixed from the radial velocity orbit of Table~\ref{tBert}.}\\
	\multicolumn{3}{l}{$^a$: Fixed according spectral type (cf. Martins 
et al. 2005).}
\end{tabular}
\end{table}

\subsection{Ephemeris}
The ASAS-3 variable star catalog \footnote {available in Internet} provides
for the $V$ data of LS~1135 an epoch $T_0 = HJD~2451872.87$ together with the
period which is used to construct the light curve as a function of the 
orbital phase. This epoch does not correspond to a time of minimum light.
We obtained the time corresponding to the principal eclipse 
with the W-D model as an output of the DC program. In this way both
photometric and radial velocity data contribute to this result. 
The epoch was selected being approximate to the time provided by ASAS. 
The ephemeris for the main minimum of the LS~1135 binary system 
then results as follows:

\begin{equation}
	Min \mbox{ }I = \rm HJD~ 2451871.911(10) + 2.\!{\!^d}7352044(20) \cdot {\it E}\\
	\label{Eph}
\end{equation}

\section{Conclusions}
We have obtained in this work a first approximation to estimate the absolute 
physical parameters of the LS~1135 O-type binary system.
From a simultaneous analysis of the radial velocities and the light curve 
with the W--D code, we find the following: 
\begin{itemize}

\item Fixing the $T_{eff}$ of the primary component to the value
corresponding to its spectral type, namely O6.5V, 
from the best fitting W--D model we find
$T_{eff}$ $25500~^{\circ}\!K$
for the secondary component. This value corresponds approximately 
to a spectral type B1V, according to the scales of different authors 
(B$\ddot{o}$hm-Vitense, 1981; Schmidt-Kaler, 1982; Crowther, 1997). 

\item The inclination of the orbital plane of the system relative to the sky 
plane was found to be $68.5 \pm 0.5 ^{\circ}$ according the best fitting 
W--D model. 
With this value of $i$, we have obtained values for the individual masses of
the primary and secondary components
$M_{1} = 30 \pm 1~ M_{\odot}$ and 
$M_{2} =  9 \pm 1~ M_{\odot}$ and for the mean radii
$R_{1} = 12 \pm 1~ R_{\odot}$ and 
$R_{2} =  5 \pm 1~ R_{\odot}$.
Both components appear to be well inside their respective Roche lobes, and
therefore LS~1135 is a detached binary still on the main sequence.
Detached eclipsing binaries with accurate elements at the high mass end of the
main sequence are exceedingly few in number, and are very much needed
as the benchmarks of theoretical stellar models. Thus LS~1135 appears as an
excellent candidate for an accurate determination of physical parameters
in a high mass binary system.
Within the uncertainties, the first values of physical parameters of the 
binary components of LS~1135 which we have determined are in fair agreement 
with recent tabulations of Galactic O star parameters based on models of
stellar atmospheres (cf. Martins et al. 2005). Of course, for more accurate
values a better defined light curve is needed, as well as spectra with
higher signal-to-noise ratio for a more accurate determination of the radial
velocities of the secondary component.

\item It is also  interesting to mention that the mass ratio which we have 
obtained from the radial velocities $q = M_{2}/M_{1}\sim 0.31$, if confirmed,
would be among the lowest values recorded for O type main sequence binaries. 
Therefore, the physical parameters of the primary O6.5V type component 
would be better defined in this case, since the perturbations introduced by 
the secondary component on the spectral behaviour of the primary
would be minor as compared to binary systems with similar components.

\end{itemize}

\section{Acknowledgments}
We are indebted to Dr. Nidia Morrell for the loan of 5 spectrograms and for 
reading a preliminary version of this paper.
We thank Federico Bareilles for providing the improved version of the Bertiau \& Grobben code.
We would like also to thank an anonymous referee for a careful reading of the manuscript.
This research has received partial support from IALP, CONICET, Argentina.

\label{lastpage}


\begin{thebibliography}{99}
\bibitem{BG}     Bertiau F.C. \& Grobben J., 1969, Ric. Astron. Spec. Vaticano 8, 1.
\bibitem{BV}     B$\ddot{o}$hm-Vitense, E., 1981, ARA\&A, 19, 295.
\bibitem{CNM}    Corti M., Niemela V., Morrell N., 2003, A\&A, 405, 571 (CNM).
\bibitem{Cr}     Crowther P.A., 1997, in IAU Symp. 189, Fundamental Stellar Properties: 
The Interaction between Observation and Theory, ed. T. R. Bedding et al. (Dordrecht: Kluwer), 137.
\bibitem{D}      Drilling J.S., 1991, ApJS, 76, 1033.
\bibitem{E}	 Eggleton P.P., 1983, ApJ, 268, 368.
\bibitem{H}	 H\o g E. et al., 2000, A\&A, 355, L27.
\bibitem{MSH}    Martins F., Schaerer D., Hillier D.J., 2005, A\&A, 436, 1049.
\bibitem{MV}     Moffat A.F.J. \& Vogt N., 1975, A\&AS 20, 85.
\bibitem{Pe}	 Perryman M.A.C., Lindegren L. et al., 1997, A\&A, 323, L49.
\bibitem{P}	 Pojma\'nski G., 2001, in IAU Colloquium 183, 
		 Small Telescope Astronomy on Global Scales; 
		 ed. Bohdan Paczynski, Wen-Pin Chen, and Claudia Lemme; 
		 ASP Conference Series, Vol. 246, 53.
\bibitem{P2}	 Pojma\'nski G., 2002, Acta Astron., 52, 397.
\bibitem{P3}     Pojma\'nski G., 2003, Acta Astron., 53, 341.
\bibitem{P4}	 Pr\v sa A., Zwitter T., 2005, ApJ, 628, 426. 
\bibitem{SK}	 Schmidt-Kaler Th., 1982, in Landolt-B\"ornstein, New Series Group, VI, Vol. 2b, ed. K. Schaifers, \& H.H.Voigt (Berlin: Springer-Verlag), 1.
\bibitem{VH}     Van Hamme W., 1993, AJ, 106, 2096.
\bibitem{W}      Wilson R.E., 1990, ApJ, 356, 613.
\bibitem{WD}     Wilson R.E., \& Devinney E.J., 1971, ApJ, 166, 605.
\bibitem{WVH}    Wilson R.E., Van Hamme W., 2004, Computing Binary Star Observables, available at 
                 ftp://ftp.astro.ufl.edu/pub/wilson/lcdc2003/
\end{thebibliography}
\end{document}